\documentclass[
 reprint,
 amsmath,amssymb,
 aps
]{revtex4-2}

\usepackage{graphicx}
\usepackage{dcolumn}
\usepackage{bm}
\usepackage{makecell}
\usepackage[normalem]{ulem}
\usepackage{xcolor}
\usepackage{newtxtext}
\usepackage{lmodern,pdftexcmds}
\usepackage{xstring}
\usepackage{mathrsfs}
\usepackage{soul}

\DeclareMathAlphabet{\mathbfsf}{\encodingdefault}{\sfdefault}{bx}{sl}

\def\Tens#1{\IfSubStr{ABCDEFGHIJKLMNOPQRSTUVWXYZabcdefghijklmnopqrstuvwxyz}{#1}{\mathbfsf{#1}}{\bm{#1}}}
\newcommand{\figref}[1]{Fig.\,\ref{#1}}
\newcommand{\Figref}[1]{Figure\,\ref{#1}}

\newcommand{\figrefp}[2]{Fig.\,\ref{#1}\,(#2)}
\newcommand{\Figrefp}[2]{Figure\,\ref{#1}\,(#2)}

\definecolor{green}{rgb}{0,0.5,0}

\begin{document}

\preprint{APS/123-QED}

\title{Viscoelasticity reshapes the frequency response of a rotating magnetic particle}

\author{Zhiyuan \surname{Zhao}$^{1,\ast}$}
\author{Tingting \surname{Sun}$^{1}$}
\author{Han \surname{Gao}$^{2}$}
\author{Ye \surname{Xu}$^{2,3}$}
\author{Mingcheng \surname{Yang}$^{4,5,\ast}$}
\author{Masao \surname{Doi}$^{6,}$}

\email{zhaozhiyuan@zust.edu.cn}
\email{mcyang@iphy.ac.cn}
\email{doi.masao@a.mbox.nagoya-u.ac.jp}

\affiliation{
$^1$Department of Physics, Zhejiang University of Science and Technology, Hangzhou, 310023, China}
\affiliation{
$^2$Hangzhou International Innovation Institute, Beihang University, Hangzhou 311115, China}
\affiliation{
$^3$School of Mechanical Engineering and Automation, Beihang University, Beijing 102206, China}
\affiliation{
$^4$Beijing National Laboratory for Condensed Matter Physics and Laboratory of Soft Matter Physics, Institute of Physics, Chinese Academy of Sciences, Beijing 100190, China}
\affiliation{
$^5$School of Physical Sciences, University of Chinese Academy of Sciences, Beijing 100049, China}
\affiliation{
$^6$Wenzhou Institute, University of Chinese Academy of Sciences, Wenzhou, 325000, China}

\date{\today}

\begin{abstract}

A magnetic particle driven by a rotating magnetic field undergoes a transition from synchronous to asynchronous rotation at a critical driving frequency.
The asynchronous dynamics is well understood in Newtonian fluids but remains unclear in viscoelastic media. 
Here, we develop a theoretical description of the asynchronous rotation of a magnetic particle in a Jeffreys-type viscoelastic fluid. 
The particle's time-averaged angular velocity exhibits a nontrivial frequency dependence that changes from non-monotonic to monotonic as
the polymer relaxation time increases. 
This behavior is explained by the interplay among magnetic driving, viscoelastic relaxation, and frequency-dependent viscous dissipation.
We further derive an asymptotic expression that captures the non-monotonic dependence.
These results clarify how solvent and polymer contributions jointly control asynchronous rotation and provide a physical basis for guiding relevant applications in complex fluids.

\end{abstract}


\maketitle


\section{\label{sec:level1}Introduction}


Magnetic particles driven by rotating magnetic fields offer a versatile platform for studying nonequilibrium physics~\cite{soni2019odd,massana2021arrested,chen2025self,luo2025flocking} and for developing microscopic, non-interventional applications~\cite{tierno2014recent,erb2016actuating,zhou2021magnetically,wang2023reconfigurable}.
In particular, below a critical field frequency, the particle can synchronously rotate with the driving field, enabling programmable rotational actuation such as in magnetic hyperthermia~\cite{egolf2016hyperthermia,barrera2024magnetic}, microfluidic manipulation~\cite{moerland2019rotating,cai2023magnetic}, and microrobotics~\cite{xie2019reconfigurable,yang2021motion,landers2025clinically}.
Once the driving frequency exceeds the critical value, the system enters a nonlinear asynchronous regime, in which the particle undergoes back-and-forth oscillatory rotation.
This transition, together with the associated critical frequency, has been exploited as a sensitive probe to monitor medium viscosity~\cite{tokarev2013probing,chevry2013magnetic,berret2016local} and to detect microbiological agents or chemical binding events~\cite{mcnaughton2007physiochemical,sinn2011asynchronous}.


In an unbounded Newtonian fluid, the time-averaged angular velocity $\langle \omega \rangle$ of a magnetic particle driven by a magnetic field rotating at angular frequency $\Omega$ is known to be given by~\cite{helgesen1990nonlinear,mcnaughton2007physiochemical}
\begin{equation}
    \langle \omega \rangle =
    \begin{cases}
        \Omega, & \Omega \le \Omega_\mathrm{c}, \\
        \Omega - \sqrt{\Omega^2-\Omega_\mathrm{c}^2}, & \Omega > \Omega_\mathrm{c}.
    \end{cases}
    \label{eq:frequency_newtonian}
\end{equation}
Here, $\Omega_\mathrm{c}$ is the critical frequency defined by $\Omega_\mathrm{c} \equiv \mu_0 m H_0/\zeta_\mathrm{r}$, $\mu_0$ is the vacuum permeability, $\zeta_\mathrm{r}$ is the rotational friction coefficient, and $m$ and $H_0$ denote the intensity of the magnetic dipole moment and the rotating field, respectively.
Equation\,\eqref{eq:frequency_newtonian} has served as a standard reference for describing the frequency response of magnetic particles with diverse geometries in rotating or precessing magnetic fields~\cite{tierno2008controlled,frka2011dynamics,tierno2009overdamped}.


However, many fluids relevant to biomedical and soft-matter applications contain high-molecular-weight polymers or other mesoscopic structures that exhibit both viscous dissipation and elastic response under deformation~\cite{errill1969rheology,fung2013biomechanics}.
Prior work~\cite{chevry2013magnetic,loosli2016viscoelasticity} reported that in Maxwell fluids, viscoelasticity modifies the asynchronous back-and-forth rotation of a superparamagnetic nanowire while leaving its time-averaged angular velocity nearly identical to the Newtonian prediction.
The result suggests that Eq.\,\eqref{eq:frequency_newtonian} may remain valid in viscoelastic fluids, especially for extracting the critical frequency from experimental measurements.


Our recent study demonstrated that this apparent equivalence holds only when $\tau \Omega_\mathrm{c} \ll 1$~\cite{gao2025response}, where $\tau$ is the single-mode relaxation time of the polymer stress.
Beyond such a limit, increasing $\tau \Omega_\mathrm{c}$ causes a discernible enhancement of $\langle \omega \rangle$ in the asynchronous regime. 
Despite the progress, the asynchronous dynamics of magnetic particles in viscoelastic fluids remain theoretically incomplete.
Particularly, an explicit relation between $\langle \omega \rangle$ and $\Omega$ is still lacking for the asynchronous regime.
Moreover, as compared with the Maxwell model, the Jeffreys model provides a more realistic description of polymer solutions by additionally incorporating an instantaneous viscous dissipation from the solvent~\cite{barnes1989introduction}.
How this solvent-mediated dissipation affects the asynchronous rotation of a magnetic particle remains to be clarified.


In this work, we investigate the asynchronous rotation of an isolated magnetic sphere driven by a rotating magnetic field in a Jeffreys fluid.
The particle’s rotational dynamics is studied by numerically solving the time evolution equation for the particle orientation over broad ranges of the field frequency, solvent-to-solution viscosity ratio, and polymer relaxation time.
We show that $\langle \omega \rangle$ at the asynchronous regime has a complex dependence on the driving frequency, including a transition from Newtonian-like decay to a non-monotonic dependence. 
Their mechanisms are then explained in terms of a competition among magnetic driving, elastic memory, and viscous dissipation.
Finally, by focusing on the non-monotonic regime, we develop a perturbative theory and derive an analytical expression for the frequency dependence of $\langle \omega \rangle$.


\section{Methods}

\begin{figure}[tb]
	\centering
	\includegraphics[width=0.48\textwidth]{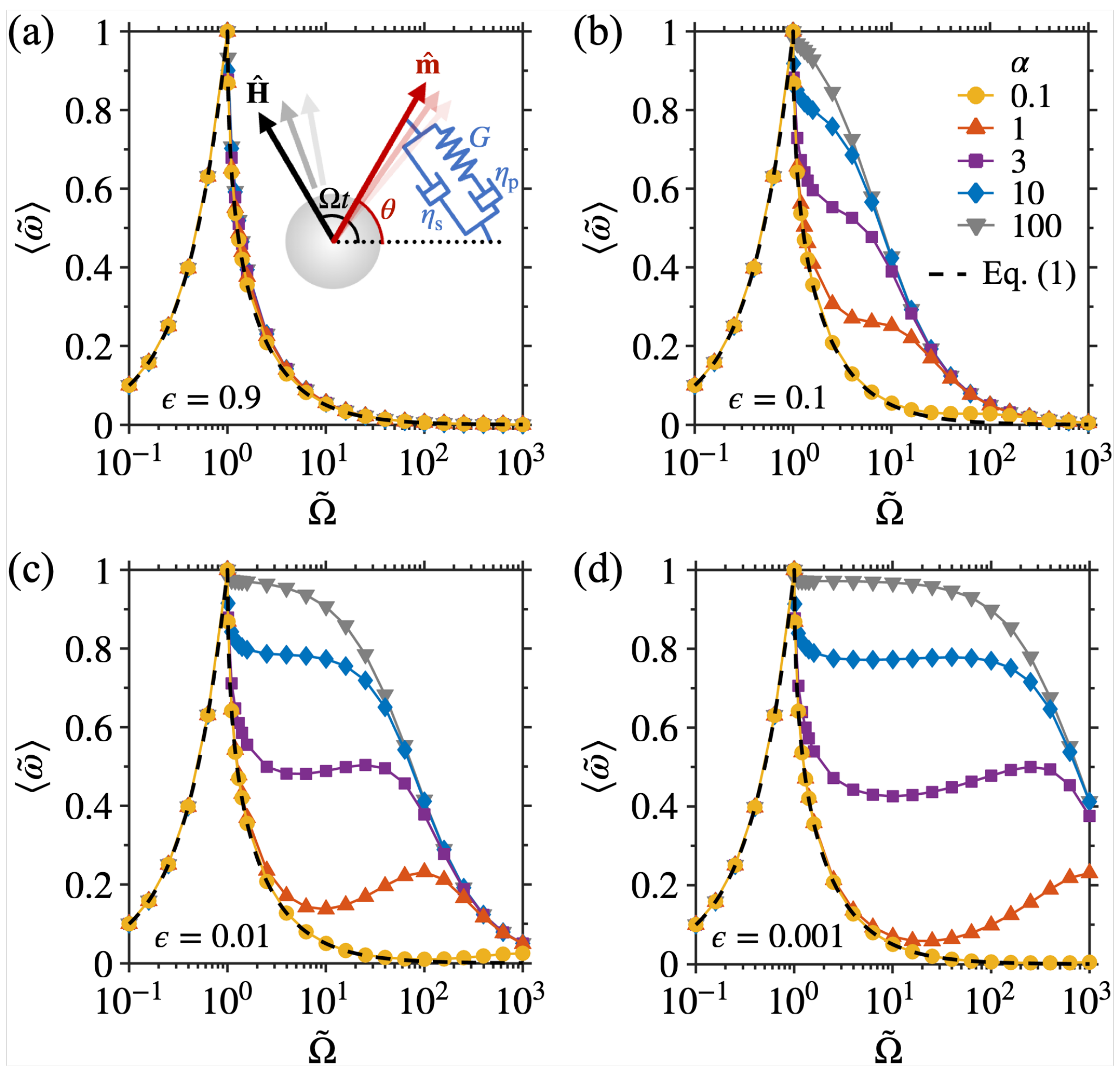}
	\caption{\label{fig:Figure_1}
    Frequency response of a rotating magnetic particle in a Jeffreys fluid.
    Dimensionless time-averaged angular velocity $\langle \tilde \omega \rangle$ as a function of scaled driving frequency $\tilde \Omega$ for (a) $\epsilon = 0.9$, (b) $0.1$, (c) $0.01$, and (d) $0.001$. 
    The legend is presented in (b).
    The black dashed curve indicates the Newtonian prediction given by Eq.\,\eqref{eq:frequency_newtonian}.
    Inset of (a): Schematic of the ferromagnetic sphere in a Jeffreys fluid.
    The vectors $\hat{\mathbf{m}}$ and $\hat{\mathbf{H}}$ denote the magnetic moment and applied field directions, whose instantaneous azimuthal angles are $\theta$ and $\Omega t$, respectively. 
    In the Jeffreys model, $\eta_\mathrm{s}$, $\eta_\mathrm{p}$, and $G$ denote the solvent viscosity, polymer viscosity, and elastic modulus, respectively. 
    }
\end{figure}

We consider a ferromagnetic sphere of radius $a$ and fixed magnetic dipole moment $m$, suspended in a Jeffreys fluid and subjected to a rotating magnetic field of amplitude $H_0$ and angular frequency $\Omega$ [\figrefp{fig:Figure_1}{a}, inset].
The particle is subjected to a magnetic torque
\begin{equation}
	T_\mathrm{m} = \mu_0 m H_0 \sin (\Omega t - \theta),
    \label{eq:magnetic_torque}
\end{equation}
where $\theta$ denotes the azimuthal angle of the particle and $t$ is time.


In the Jeffreys model~\cite{barnes1989introduction}, a Maxwell element and a Newtonian dashpot are connected in parallel, accounting for the delayed polymeric stress and the instantaneous solvent dissipation, respectively.
The corresponding elastic torque $T_\mathrm{e}$ and solvent viscous toque $T_\mathrm{v}$ acting on the particle are given by
\begin{gather}
    T_\mathrm{e} = -8 \pi a^3 \int_{-\infty}^t G(t-t') \dot{\theta}(t') dt',
    \label{eq:elastic_torque} \\
    T_\mathrm{v} = -8 \pi a^3 \eta_\mathrm{s} \dot{\theta},
    \label{eq:viscous_torque}
\end{gather}
where an overdot denotes derivative with respect to time, $\dot \theta$ represents the instantaneous particle angular velocity (equivalent to  $\omega$), and $\eta_\mathrm{s}$ is the solvent viscosity.
The polymer relaxation modulus is
\begin{equation}
	G(t) = \sum_i G_i e^{-t/\tau_i},
    \label{eq:D3}
\end{equation}
where $G_i$ and $\tau_i$ are the elastic modulus and relaxation time associated with the $i$th relaxation mode, respectively. 


Neglecting rotational inertial and thermal fluctuations, the deterministic torque balance on the particle is
\begin{equation}
	T_\mathrm{m} + T_\mathrm{e} + T_\mathrm{v} = 0.
    \label{eq:torque_balance_jeffreys}
\end{equation}
By considering a single relaxation mode and by defining the phase lag between the field and the particle as $\phi \equiv \Omega t - \theta$, we rewrite Eq.\,\eqref{eq:torque_balance_jeffreys} to
\begin{equation}
    \Omega - \dot \phi
    =
    -\frac{\eta_\mathrm{p}}{\eta_\mathrm{s} \tau} \theta_{\mathrm{e}} 
    +
    \Omega_\mathrm{c} \sin \phi,
    \label{eq:jeffreys_eq_1}
\end{equation}
where $\eta_\mathrm{p}$ and $\tau$ denote the polymer viscosity and the single-mode relaxation time, respectively.
The auxiliary variable $\theta_{\mathrm{e}}$ represents the elastic rotational memory stored in the polymer mode and satisfies
\begin{gather}
    \theta_\mathrm{e}
    = 
    \int^t_{-\infty} 
    e^{-(t-t')/\tau} \dot \theta(t') dt',
    \label{eq:theta_e} \\
    \dot \theta_{\mathrm{e}}
    = 
    - \frac{\theta_{\mathrm{e}}}{\tau} 
    + 
    \Omega - \dot \phi.
    \label{eq:jeffreys_eq_2}
\end{gather}


We introduce the dimensionless variables
\begin{equation}
    \epsilon = \frac{\eta_\mathrm{s}}{\eta_\mathrm{tot}}, \quad
    \alpha = \tau \Omega^\ast_\mathrm{c}, \quad
    \tilde \Omega = \frac{\Omega}{\Omega^\ast_\mathrm{c}}, \quad
    \tilde t = t \Omega^\ast_\mathrm{c},
    \label{eq:nondimensionalization}
\end{equation}
where $\eta_\mathrm{tot} \equiv \eta_\mathrm{s} + \eta_\mathrm{p}$ is the zero-shear viscosity and 
\begin{equation}
   \Omega^\ast_\mathrm{c}
   \equiv
   \frac{\mu_0 m H_0}{8 \pi a^3 \eta_\mathrm{tot}}
   \label{eq:jeffreys_omega_c}
\end{equation}
is the corresponding critical frequency.
The solvent-to-solution viscosity ratio $\epsilon \in (0,1)$ approaches the Newtonian limit as $\epsilon \rightarrow 1$ and the Maxwell limit as $\epsilon \rightarrow 0$.
Hereafter, the overdots denote derivatives with respect to $\tilde t$.
Then, Eqs.\,\eqref{eq:jeffreys_eq_1} and \eqref{eq:jeffreys_eq_2} become
\begin{gather}
    \dot \phi
    = 
    \frac{1}{\epsilon} \left( 
    \frac{1 - \epsilon}{\alpha} \theta_\mathrm{e}
    -
    \sin \phi
    \right)
    +
    \tilde \Omega,
    \label{eq:jeffreys_eq_1s} \\
    \dot \theta_\mathrm{e}
    = 
    \frac{1}{\epsilon} \left( 
    \sin \phi - \frac{\theta_\mathrm{e}}{\alpha} 
    \right).
    \label{eq:jeffreys_eq_2s}
\end{gather}


We note that although the derivation above assumes a ferromagnetic sphere, Eqs.\,\eqref{eq:jeffreys_eq_1s} and \eqref{eq:jeffreys_eq_2s} can be extended to other overdamped magnetic particles with distinct geometries and magnetic responses.
Specifically, the particle geometry enters through the rotational friction coefficient, whereas the magnetic properties enter through the magnitude of Eq.\,\eqref{eq:magnetic_torque}.
These changes alter only the definition of $\Omega^\ast_\mathrm{c}$. 
In the rapid-relaxation limit, $\alpha \ll 1$, Eqs.\,\eqref{eq:jeffreys_eq_1s} and \eqref{eq:jeffreys_eq_2s} reduce to~\cite{gao2025response}
\begin{equation}
    \tilde \Omega - \dot \phi =  \sin \phi,
    \label{eq:jeffreys_phi_approx}
\end{equation}
whose solution gives the Newtonian response in Eq.\,\eqref{eq:frequency_newtonian} but with $\Omega_\mathrm{c}$ replaced by $\Omega^\ast_\mathrm{c}$.


We solve Eqs.\,\eqref{eq:jeffreys_eq_1s} and \eqref{eq:jeffreys_eq_2s} numerically using a fourth-order Runge-Kutta scheme over $\epsilon \in [0.001,0.9]$, $\alpha \in [0.1,100]$, and $\tilde \Omega \in [0.1,1000]$.
The minimum time step is set to $\Delta \tilde t = 2\pi/(10^5 \tilde \Omega)$, corresponding to $10^5$ integration steps per field period. 
Each trajectory is integrated until the absolute difference of net advances of $\theta$ between two successive cycles falls below $10^{-5}$, at which point the system is considered to have reached a steady oscillatory state.
Then, the dimensionless time-averaged angular velocity, $\langle \tilde{\omega} \rangle \equiv \langle \omega \rangle / \Omega^\ast_\mathrm{c}$, is determined from the slope of a linear fit to $\theta(\tilde t)$ over a steady-state sampling window spanning approximately $300$ field periods.
We have verified that the results are insensitive to further reductions in the time step and to further broaden the sampling window for averaging.


\section{Results}


\subsection{Frequency dependence of $\langle \tilde \omega \rangle$}


\Figref{fig:Figure_1} shows the dependence of $\langle \tilde \omega \rangle$ on $\tilde \Omega$ for various $\alpha$ and $\epsilon$. 
For $\tilde \Omega \le 1$, all curves collapse onto the Newtonian prediction (black dashed line) with $\langle \tilde \omega \rangle = \tilde \Omega$, indicating phase-locked synchronous rotation of the magnetic particle.
As reported in our recent work~\cite{gao2025response}, in such a frequency regime, the system relaxes to a stable fixed point characterized by $\phi = \arcsin \tilde \Omega$ and $\theta_\mathrm{e} = \alpha$. 
Variations in $\epsilon$ and $\alpha$ merely shift the location of the fixed point but do not alter the phase-locked nature of the particle motion.


For $\tilde \Omega > 1$, the phase lag is no longer stationary, and the particle enters an asynchronous phase-slipping state. 
At $\epsilon = 0.9$ that approximates the Newtonian limit, the response remains close to the Newtonian prediction and depends only weakly on $\alpha$ [\figrefp{fig:Figure_1}{a}]. 
By contrast, as $\epsilon$ decreases towards the Maxwell limit, $\langle \tilde \omega \rangle$ displays a complex dependence on $\tilde \Omega$, which is increasingly deviated from the Newtonian prediction and becomes strongly dependent on $\alpha$ [\figrefp{fig:Figure_1}{b} to (d)].


Particularly, for sufficiently short relaxation times, $\langle \tilde \omega \rangle$ first decreases with $\tilde \Omega$, then increases over an intermediate frequency interval, and finally returns to a decreasing high-frequency branch. 
Similar non-monotonic responses have been observed experimentally~\cite{haapanen2015observation,loosli2016viscoelasticity}, although their origin was not explicitly resolved.
Furthermore, \figref{fig:Figure_1} shows that increasing $\alpha$ enhances $\langle \tilde \omega \rangle$ and eventually merges the two extrema into a broad shoulder, producing a monotonic response. 
Decreasing $\epsilon$ not only increases $\langle \tilde \omega \rangle$ in the high-frequency limit, but also broadens the frequency interval over which the non-monotonicity or shoulder occurs. 


\subsection{Memory-mediated intracycle dynamics}

\begin{figure}[tb]
	\centering
	\includegraphics[width=0.48\textwidth]{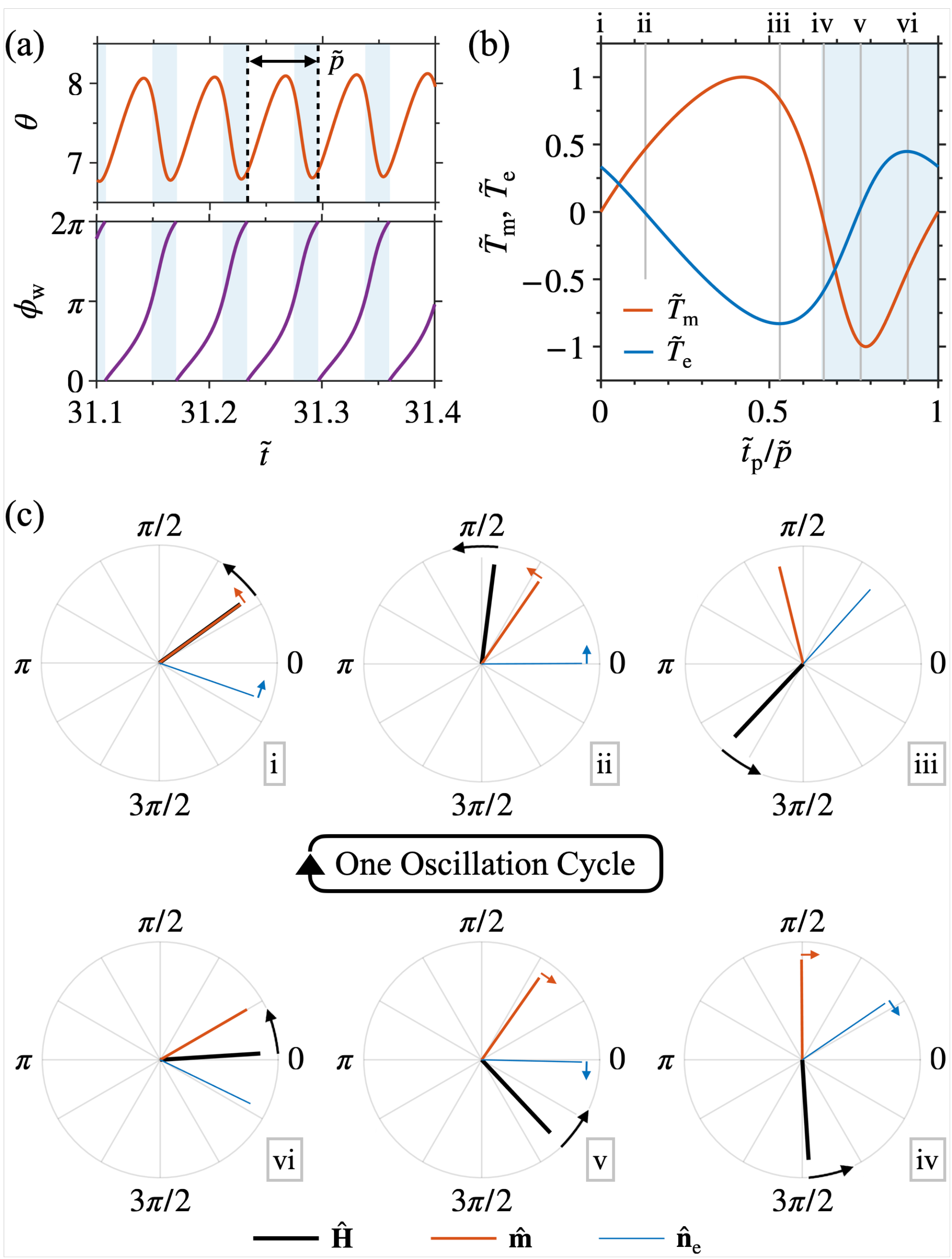}
	\caption{ \label{fig:Figure_2}
    Memory-mediated dynamics within one phase-slip cycle.
    (a) Steady-state time evolution of the particle orientation $\theta$ and the wrapped phase lag $\phi_\mathrm{w}$. 
    Shaded and unshaded intervals correspond to $\phi_\mathrm{w} \in [\pi,2\pi)$ and $\phi_\mathrm{w} \in [0,\pi)$, respectively. 
    Dashed lines delimit one dimensionless oscillation period $\tilde p$.
    (b) Dimensionless magnetic torque $\tilde T_\mathrm{m}$ and elastic torque $\tilde T_\mathrm{e}$ as a function of normalized intracycle time $\tilde t_{\mathrm p}$. 
    (c) Relative angular configurations of the magnetic field $\hat{\mathbf{H}}$, the magnetic moment $\hat{\mathbf{m}}$, and the auxiliary elastic-deformation vector $\hat{\mathbf{n}}_\mathrm{e}$ at the six representative instants. 
    The azimuthal angles of these vectors are $\Omega t$, $\theta$, and $\theta_\mathrm{e}$, respectively. 
    The state $\theta_\mathrm{e} = 0$ corresponds to a relaxed Maxwell branch. 
    Curved black, red, and blue arrows indicate the instantaneous evolution directions of $\Omega t$, $\theta$, and $\theta_\mathrm{e}$, respectively. 
    For the vectors without accompanying curved arrows, they are in the states with zero changing rate.
    All panels show $\epsilon = 0.01$, $\alpha = 1$, and $\tilde \Omega = 100$.
    }
\end{figure}

To identify the dynamical origin of the complex frequency dependence of $\langle\tilde\omega\rangle$, we examine a representative asynchronous state with $\epsilon = 0.01$, $\alpha = 1$, and $\tilde\Omega = 100$.
\Figrefp{fig:Figure_2}{a} manifests the steady-state evolution of the particle angle $\theta$ and the wrapped phase lag $\phi_{\mathrm w} \equiv \phi \bmod 2\pi$.
We define the oscillation period $p$ as the time required for $\phi$ to increase by $2\pi$, with the dimensionless form given by $\tilde p = p \Omega_{\mathrm c}^{\ast}$.
During one cycle, the particle first undergoes a forward excursion, then a backward recoil, and finally a short forward rotation.
The shaded and unshaded intervals correspond to $\phi_{\mathrm w} \in [\pi,2\pi)$ and $\phi_{\mathrm w} \in [0,\pi)$, respectively.


The intracycle torque evolution is shown in \figrefp{fig:Figure_2}{b}, where $\tilde T_{\mathrm m} \equiv T_{\mathrm m} / (\mu_0 m H_0)$, $\tilde T_{\mathrm e} \equiv T_{\mathrm e} / (\mu_0 m H_0)$, and $\tilde t_{\mathrm p} \in [0,\tilde p)$ denotes the normalized intracycle time.
The offset of the extrema and zero crossings between the two curves reflects the delayed elastic response to particle rotation.
Because the rotational dynamics are overdamped, the sign of $\tilde T_{\mathrm m} + \tilde T_{\mathrm e}$ determines the instantaneous direction of particle rotation.
Six representative instants are selected to illustrate the intracycle evolution of the relative angular configurations of the magnetic field ($\hat{\mathbf H}$), magnetic moment ($\hat{\mathbf m}$), and elastic memory ($\hat{\mathbf n}_{\mathrm e}$) [\figrefp{fig:Figure_2}{c}].
Notice that the vector $\hat{\mathbf n}_\mathrm{e}$ is a visualization aid rather than an additional physical degree of freedom; it marks $\theta_\mathrm{e}$ relative to zero deformation.


At the beginning of the cycle (instant i), the magnetic moment is aligned with the field and $\tilde T_{\mathrm m} = 0$.
However, residual negative deformation from the preceding cycle generates a positive elastic torque that drives the particle forward (in the direction of the field rotation).
As the field advances, the elastic deformation changes sign at instant ii, causing $\tilde T_{\mathrm e}$ to become negative and oppose the forward rotation.
At instant iii, the magnetic and elastic torques balance, marking the forward turning point where $\theta$ reaches a local maximum and the particle begins to backward recoil.


By instant iv, $\tilde T_{\mathrm m}$ becomes negative and acts together with the negative elastic torque to drive the backward rotation.
Under the persistent negative magnetic torque, the elastic deformation crosses zero at instant v and then accumulates with the opposite sign.
At instant vi, the positive elastic torque balances the negative magnetic torque; $\theta$ reaches a local minimum, defining the backward turning point.
As $\phi_{\mathrm w}$ approaches $2\pi$, the magnitude of the negative magnetic torque decreases, whereas the positive elastic torque persists because of the finite relaxation time.
Consequently, $\tilde T_{\mathrm m}+\tilde T_{\mathrm e}$ remains transiently positive, driving the particle to rotate forward and briefly lead the field.


This sequence demonstrates that elastic memory reshapes the net torque within each oscillation cycle and breaks the dynamical equivalence of the forward and backward portions. 
Especially, deformation retained from one cycle affects the next.
This is different from the asynchronous dynamics in a Newtonian fluid, which is governed solely by the instantaneous magnetic torque.
For the Newtonian case, the particle rotates forward for $0 < \phi_{\mathrm w} < \pi$ and backward for $\pi < \phi_{\mathrm w} < 2\pi$, with turning points occurring when the magnetic torque vanishes.
The resulting phase-slip rate relies only on the driving frequency, giving rise to the monotonic frequency response described by Eq.\,\eqref{eq:frequency_newtonian}.


\subsection{Origin of the non-monotonic response}

\begin{figure}[tb]
	\centering
	\includegraphics[width=0.48\textwidth]{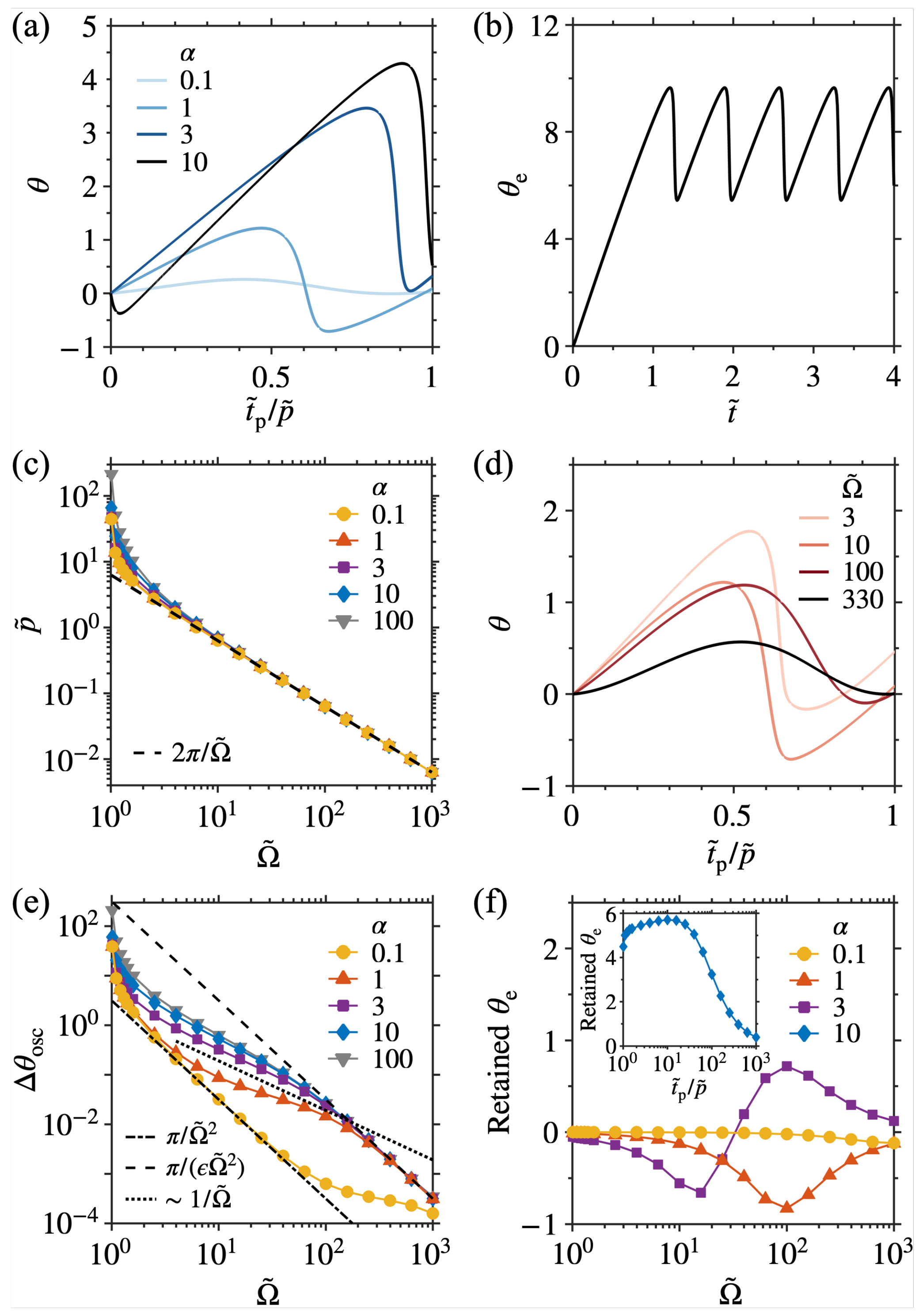}
	\caption{ \label{fig:Figure_3}
    Intracycle origin and scaling of the non-monotonic response.
    Intracycle evolution of $\theta$ (a) at $\tilde\Omega = 10$ for various $\alpha$ and (d) at $\alpha = 1$ for various $\tilde\Omega$.
    (b) Time evolution of $\theta_\mathrm{e}$ for $\tilde\Omega = 10$ and $\alpha = 10$.
    Frequency dependence of (c) the oscillation period $\tilde p$ and (e) the net angular advance per cycle $\Delta \theta_{\mathrm{osc}}$ for different $\alpha$.
    The dashed line in (c) denotes the period of the rotating field.
    In (e), the dashed and dot-dash lines denote $\Delta \theta_\mathrm{osc} = \pi / \tilde \Omega^2$ and $\Delta \theta_\mathrm{osc} = \pi / (\epsilon \tilde \Omega^2)$, respectively. 
    The dotted line indicates power law with exponent $-1$.
    (f) Frequency dependence of the retained elastic deformation for various $\alpha$.
    All results are obtained for $\epsilon=0.01$.
    }
\end{figure}

\Figrefp{fig:Figure_3}{a} shows the evolution of $\theta$ against the normalized intracycle time at $\tilde \Omega = 10$ and $\epsilon = 0.01$.
As $\alpha$ increases, both the angular displacement and the duration of the forward rotation become much larger than those of the backward rotation, leading to a increase of the net angular advance per cycle, $\Delta \theta_\mathrm{osc}$. 
This phenomenon can be explained by fact that a longer relaxation time promotes accumulation and persistence of elastic stress.
The delayed buildup of this stress extends the forward excursion and meanwhile accelerates the backward recoil.
However, for $\alpha = 10$, the elastic load becomes so strong that $\theta_\mathrm{e}$ initially exceeds $2\pi$ [\figrefp{fig:Figure_3}{b}].
It can no longer fully relax during the subsequent oscillations, but instead reaches a steady oscillatory state characterized by a finite snap-back angle less than $2\pi$.
The resulting positive retained deformation leads to absent elasticity-assisted forward rotation and short recoiling of the particle at the beginning of each cycle. 


In the steady asynchronous case, the time-averaged angular velocity can equivalently be estimated by
\begin{equation}
    \langle \tilde{\omega} \rangle 
    =
    \frac{ \Delta \theta_\mathrm{osc} }{ \tilde p }.
    \label{eq:dtheta_to_p}
\end{equation}
As observed in \figrefp{fig:Figure_3}{c}, for $\tilde \Omega \rightarrow 1^+$, $\tilde p$ exceeds the driving period (black dashed line) by several orders of magnitude. 
This is understood by critical slowing down of the phase-slip dynamics near the synchronization threshold.
Away from this threshold, $\tilde p$ asymptotically approaches the driving period and becomes independent of $\alpha$.
In this case, the frequency dependence of $\langle \tilde \omega \rangle$ is governed primarily by $\Delta \theta_\mathrm{osc}$.
\Figrefp{fig:Figure_3}{a} exemplifies that increasing $\alpha$ at $\tilde \Omega = 10$ gives rise to an enhancement of $\langle \tilde \omega \rangle$. 


The influence of $\tilde \Omega$ on the intracycle evolution of $\theta$ is illustrated in \figrefp{fig:Figure_3}{d} for $\alpha = 1$ and $\epsilon = 0.01$.
Because a faster rotating field prevents sustained forward rotation yet diminishes elastic effect accelerating the recoil process, it attenuates the oscillatory asymmetry and decreases $\Delta \theta_\mathrm{osc}$. 
\Figrefp{fig:Figure_3}{e} exhibits that at both low-frequency and high-frequency limits, curves for different $\alpha$ collapse into the asymptotic power laws proportional to $\tilde \Omega^{-2}$.
However, at the intermediate-frequency regime, the scaling exponent increases and even exceeds $-1$ for $\alpha \le 1$.
This is because the backward excursion is suppressed more rapidly than the forward excursion as $\tilde \Omega$ increases [\figrefp{fig:Figure_3}{d}].
Since $\tilde p$ scales $\tilde \Omega^{-1}$ away from the threshold, $\Delta \theta_\mathrm{osc}$ with a scaling exponent larger than $-1$ eventually produces a local increase in $\langle \tilde \omega \rangle$ with driving frequency.


In the steady asynchronous state, the back-and-forth particle rotation imposes an oscillatory deformation on the surrounding medium. 
For the Jeffreys fluid, its complex viscosity reads~\cite{ponce2018validity}
\begin{equation}
    \eta^\ast(\Omega_\phi)
    =
    \eta_\mathrm{tot} 
    \frac{ 1 + i \epsilon \Omega_\phi \tau }
    { 1 + i \Omega_\phi \tau },
    \label{eq:complex_viscosity}
\end{equation}
where the angular frequency of the particle oscillation $\Omega_\phi \equiv 2\pi / p$.
The real part, or the dynamic viscosity, is then given by
\begin{equation}
    \operatorname{Re} [\eta^\ast(\Omega_\phi)]
    =
    \eta_\mathrm{tot} 
    \frac{1+\epsilon(\Omega_\phi \tau)^2}
    {1+(\Omega_\phi \tau)^2}.
    \label{eq:dynamic_viscosity}
\end{equation}
It reduces from $\eta_\mathrm{tot}$ to $\eta_\mathrm{s}$ as $\Omega_\phi$ increases, reflecting a crossover from polymer-assisted dissipation to solvent-dominated dissipation.
This transition is controlled by two timescales: the relaxation time $\tau$ and the retardation time $\epsilon \tau$, the latter of which quantifies the lag of polymer deformation behind the applied stress.
By comparing these material timescales with $p$, we determine three stages in the frequency response of $\langle \tilde \omega \rangle$. 


Stage I corresponds to $p > 2 \pi \tau$, where the polymeric deformation relaxes almost completely within each cycle, so that little elastic memory is transferred to the following cycle.
In this stage, elastic memory has a weak influence on the particle, leading the frequency response of $\langle \tilde \omega \rangle$ to approach the Newtonian prediction based on $\eta_\mathrm{tot}$.
Away from the synchronous threshold, such a prediction can be approximated to 
\begin{equation}
    \langle \tilde \omega \rangle
    \simeq
    \frac{1}{2 \tilde \Omega}.
    \label{eq:omega_avg_approx_1}
\end{equation}
By applying $\tilde p \simeq 2\pi/\tilde \Omega$, we eventually obtain $\Delta \theta_\mathrm{osc} \simeq \pi / {\tilde \Omega}^2$.


Stage II is defined by $2 \pi \epsilon \tau < p < 2\pi \tau$.
Within this interval, polymeric stress relaxes only partially during each cycle, negatively enhancing the retained elastic deformation [\figrefp{fig:Figure_3}{f}].
The resulting elasticity-assisted forward rotation can produce a slower decrease of $\Delta \theta_\mathrm{osc}$ than $\tilde \Omega^{-1}$ and then a positive dependence of $\langle \tilde \omega \rangle$ on $\tilde \Omega$.
However, when the retained elastic deformation becomes positive, the elasticity-assisted forward rotation is absent.
This leads to $\Delta \theta_\mathrm{osc} \sim \tilde \Omega^{-1}$ and then a transient plateau in the frequency response of $\langle \tilde \omega \rangle$.


Stage III corresponds to $p < 2 \pi \epsilon \tau$, where the polymeric memory changes weakly during a individual cycle, even though a finite deformation may persist across successive cycles.
The solvent branch becomes the dominant dissipative pathway, so that $\langle \tilde \omega \rangle$ is again independent of $\alpha$ and decreases monotonically with increasing $\tilde \Omega$.
Numerically, $\Delta \theta_\mathrm{osc}$ approaches $\pi / (\epsilon \tilde \Omega^2)$, exhibiting a distinct prefactor from the asymptotic expression for Stage I. 
We will derive and explain it in the next subsection.


\subsection{Asymptotic solution of $\langle \tilde \omega \rangle$ in the non-monotonic regime}


To obtain an analytical expression for the non-monotonic regime, we rewrite the torque balance as
\begin{equation}
    \mathcal{G}[\phi(t)]
    \equiv
    \mathcal{G}_1[\phi(t)] - \Omega + \mathcal{G}_2[\phi(t)],
    \label{eq:G}
\end{equation}
where 
\begin{gather}
    \mathcal{G}_1[\phi(t)]
    \equiv
    \Omega^\ast_\mathrm{c} \sin \phi,
    \label{eq:G1}
    \\
    \mathcal{G}_2[\phi(t)]
    \equiv
    \epsilon \dot \phi
    +
    (1 - \epsilon) \int^t_{-\infty} \frac{dt'}{\tau} 
    \mathrm{e}^{-(t-t')/\tau} \dot \phi(t').
    \label{eq:G2}
\end{gather}


Considering the steady-state phase lag satisfies $\phi_\mathrm{st}(t+p) = \phi_\mathrm{st}(t) + 2 \pi$, we approximate it using a first-order Fourier expansion
\begin{equation}
    \phi_\mathrm{st}(t)
    =
    \Omega_\phi t
    +
    A \cos(\Omega_\phi t)
    + 
    B \sin(\Omega_\phi t),
    \label{eq:phi_st}
\end{equation}
where $A$ and $B$ are the small perturbation coefficients.
Because $\mathcal{G}[\phi(t)] = 0$, its cycle average satisfies
\begin{gather}
    \langle \mathcal{G}[\phi_\mathrm{st}(t)] \rangle 
    = 0,
    \\
    \langle \mathcal{G}[\phi_\mathrm{st}(t)] \cos(\Omega_\phi t) \rangle 
    = 0,
    \\
    \langle \mathcal{G}[\phi_\mathrm{st}(t)] \sin(\Omega_\phi t) \rangle
    = 0.
\end{gather}


By substituting Eqs.\,\eqref{eq:phi_st} into \eqref{eq:G1}, we obtain
\begin{align}
    \mathcal{G}_1[\phi_\mathrm{st}(t)]
    & \approx \notag
    \Omega^\ast_\mathrm{c} \sin (\Omega_\phi t) \\
    & + \Omega^\ast_\mathrm{c} \cos (\Omega_\phi t)
    [A \cos (\Omega_\phi t) + B \sin (\Omega_\phi t)].
    \label{eq:G1_expansion}
\end{align}
Its cycle average satisfies
\begin{gather}
    \langle \mathcal{G}_1[\phi_\mathrm{st}(t)] \rangle 
    = 
    \frac{\Omega^\ast_\mathrm{c}}{2} A,
    \\
    \langle \mathcal{G}_1[\phi_\mathrm{st}(t)] \cos(\Omega_\phi t) \rangle 
    = 0,
    \\
    \langle \mathcal{G}_1[\phi_\mathrm{st}(t)] \sin(\Omega_\phi t) \rangle
    = \frac{\Omega^\ast_\mathrm{c}}{2}.
\end{gather}


Similarly, substituting Eqs.\,\eqref{eq:phi_st} into \eqref{eq:G2} yields
\begin{equation}
    \mathcal{G}_2[\phi_\mathrm{st}(t)]
    =
    \Omega_\phi [1 + C \cos (\Omega_\phi t) + D \sin (\Omega_\phi t)],
    \label{eq:G2_expansion}
\end{equation}
with 
\begin{gather}
    C = \epsilon B + (1 - \epsilon) \frac{A \Omega_\phi \tau + B}{1 + (\Omega_\phi \tau)^2},
    \\
    D = -\epsilon A - (1 - \epsilon) \frac{A - B \Omega_\phi \tau}{1 + (\Omega_\phi \tau)^2}.
\end{gather}
The corresponding cycle average satisfies
\begin{gather}
    \langle \mathcal{G}_2[\phi_\mathrm{st}(t)] \rangle = \Omega_\phi,
    \\
    \langle \mathcal{G}_2[\phi_\mathrm{st}(t)] \cos(\Omega_\phi t) \rangle 
    = \frac{\Omega_\phi}{2} C,
    \\
    \langle \mathcal{G}_2[\phi_\mathrm{st}(t)] \sin(\Omega_\phi t) \rangle
    = \frac{\Omega_\phi}{2} D.
\end{gather}


By projecting Eq.\,\eqref{eq:G} onto the first harmonic, we have
\begin{gather}
    0
    =
    \langle  \mathcal{G}_1 \cos (\Omega_\phi t)\rangle
    +
    \langle  \mathcal{G}_2 \cos (\Omega_\phi t)\rangle,
    \label{eq:Gcos_relation}
    \\
    0
    =
    \langle  \mathcal{G}_1 \sin (\Omega_\phi t)\rangle
    +
    \langle  \mathcal{G}_2 \sin (\Omega_\phi t)\rangle.
    \label{eq:Gsin_relation}
\end{gather}
Solving Eqs.\,\eqref{eq:G}, \eqref{eq:Gcos_relation}, and \eqref{eq:Gsin_relation} gives
\begin{gather}
    \Omega - \Omega_\phi
    =
    \frac{A}{2} \Omega^\ast_\mathrm{c},
    \label{eq:omega_avg}
    \\
    A =
    \frac{\Omega^\ast_\mathrm{c}}{\Omega_\phi} 
    \frac{1 + \epsilon \Omega^2_\phi \tau^2}
    {1 + \epsilon^2 \Omega^2_\phi \tau^2},
    \label{eq:A}
    \\
    B =
    -\frac{(1 - \epsilon) \tau \Omega^\ast_\mathrm{c}}
    {1 + \epsilon^2 \Omega^2_\phi \tau^2}.
    \label{eq:B}
\end{gather}


Because $\langle \omega \rangle = \Omega - \Omega_\phi$, Eq.\,\eqref{eq:omega_avg} leads to
\begin{equation}
    \langle \tilde \omega \rangle  
    =
    \frac{1}{2 \tilde \Omega_\phi} 
    \frac{1 + \epsilon {\tilde \Omega}^2_\phi \alpha^2}
    {1 + \epsilon^2 {\tilde \Omega}^2_\phi \alpha^2}, 
    \label{eq:omega_avg_analytical_original}
\end{equation}
where $\tilde \Omega_\phi = \Omega_\phi / \Omega^\ast_\mathrm{c}$. 
For the regime with $\tilde \Omega_\phi \simeq \tilde \Omega$, Eq.\,\eqref{eq:omega_avg_analytical_original} reduces to
\begin{equation}
    \langle \tilde \omega \rangle  
    =
    \frac{1}{2 \tilde \Omega} 
    \frac{1 + \epsilon {\tilde \Omega}^2 \alpha^2}
    {1 + \epsilon^2 {\tilde \Omega}^2 \alpha^2}
    =
    \frac{1}{2 \tilde \Omega} 
    \operatorname{Re} \left[ \frac{\eta_\mathrm{tot}}{\eta^\ast(\Omega)} \right],
    \label{eq:omega_avg_analytical}
\end{equation}
where $1/(2\tilde {\Omega})$ denotes the high-frequency Newtonian asymptote, while $\operatorname{Re}[\eta_{\rm tot}/\eta^*(\Omega)]$ quantifies the viscoelastic enhancement of the oscillatory mobility.


Equation\,\eqref{eq:omega_avg_analytical} shows the asymptotic solution of $\langle \tilde \omega \rangle$ in the non-monotonic regime.
It is valid only when the system is
far above the synchronous threshold and $\max(|A|,|B|) \ll 1$. 
In the limit of $\epsilon \Omega \tau \ll 1$, Eq.\,\eqref{eq:omega_avg_analytical} recovers Eq.\,\eqref{eq:omega_avg_approx_1}, whereas for $\epsilon \Omega \tau \gg 1$ it approaches
\begin{equation}
    \langle \tilde \omega \rangle
    \simeq
    \frac{1}{2 \epsilon \tilde \Omega}.
    \label{eq:omega_avg_approx_2}
\end{equation}
Substituting $\tilde p \simeq 2 \pi / \tilde \Omega$ into Eq.\,\eqref{eq:omega_avg_approx_2} gives $\Delta \theta_\mathrm{osc} \simeq \pi/(\epsilon \tilde \Omega^2)$, which is consistent with the numerical limit for Stage III [\figrefp{fig:Figure_3}{d}].


\begin{figure}[tb]
    \centering
    \includegraphics[width=0.48\textwidth]{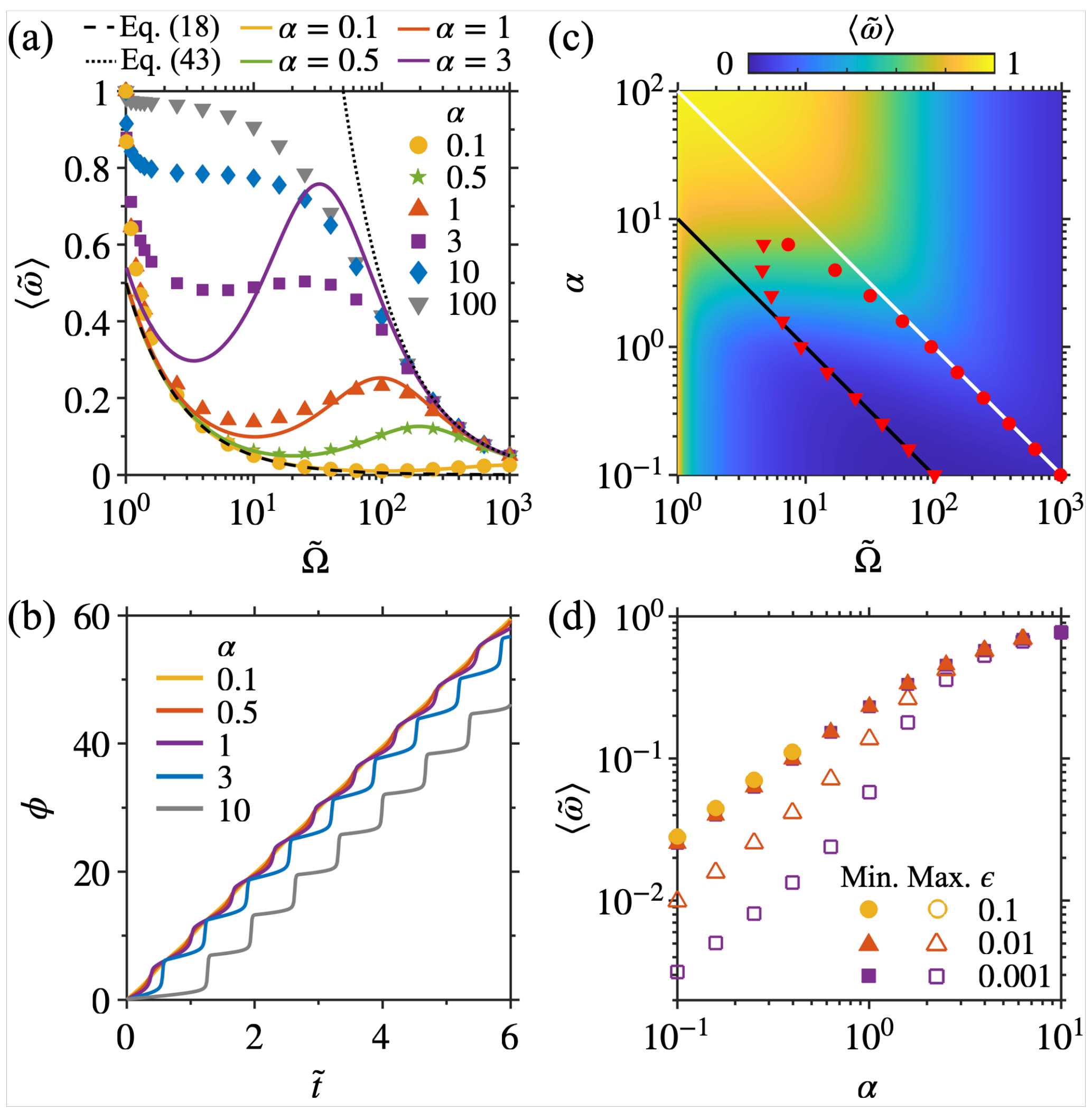}
    \caption{\label{fig:Figure_4}
    Perturbative theory for the non-monotonic response.
    (a) Frequency response of $\langle \omega \rangle$ at $\epsilon = 0.01$ and for various $\alpha$. 
    Symbols and solid lines denote the numerical result and the asymptotic solution in Eq.\,\eqref{eq:omega_avg_analytical}, respectively. 
    Dashed and dotted curves show Eqs.\,\eqref{eq:omega_avg_approx_1} and \eqref{eq:omega_avg_approx_2}, respectively. 
    (b) Numerical result of time evolution of $\phi$ for $\epsilon = 0.01$, $\tilde \Omega = 10$, and various $\alpha$.
    (c) Map of $\langle \tilde \omega \rangle$ in the plane of $\tilde \Omega$ and $\alpha$ for $\epsilon = 0.01$ from numerical simulations. 
    Solid triangles and circles represent numerical local minima and maxima, respectively.
    Black and the white solid lines show Eqs.\,\eqref{eq:omega_avg_valley} and \eqref{eq:omega_avg_peak}, respectively. 
    (d) Values of the local minima and maxima as a function of $\alpha$ for various $\epsilon$. 
    }
\end{figure}


\Figrefp{fig:Figure_4}{a} shows that Eq.\,\eqref{eq:omega_avg_analytical} agrees with the numerical results for $\alpha \le 1$ and captures the low- and high-frequency limits in Eqs.\,\eqref{eq:omega_avg_approx_1} and \eqref{eq:omega_avg_approx_2}.
The approximation deteriorates for $\alpha \ge 3$, because the assumption of small harmonic in Eq.\,\eqref{eq:phi_st} fails.
As shown in \figrefp{fig:Figure_4}{b}, numerically-solved $\phi$ becomes step-like for $\alpha \ge 3$, corresponding to the dynamics that the particle aligns with the rotating field for extended intervals and then rapidly slips backward. 


By setting $d\langle \tilde \omega \rangle / d\tilde \Omega = 0$ in Eq\,\eqref{eq:omega_avg_analytical}, we obtain the field frequencies corresponding to the local minimum and maximum:
\begin{gather}
    \tilde \Omega_\mathrm{min}
    =
    \frac{1}{\epsilon \alpha}
    \sqrt{\frac{ 1 - 3 \epsilon 
    -
    \sqrt{1 - 10 \epsilon + 9 \epsilon^2 } }{2}},
    \label{eq:omega_avg_valley}
    \\
    \tilde \Omega_\mathrm{max}
    =
    \frac{1}{\epsilon \alpha}
    \sqrt{\frac{ 1 - 3 \epsilon 
    +
    \sqrt{1 - 10 \epsilon + 9 \epsilon^2 } }{2}}.
    \label{eq:omega_avg_peak}
\end{gather} 
Real and distinct extrema additionally require $\epsilon \le 1/9$.
\Figrefp{fig:Figure_4}{c} maps $\langle \tilde \omega \rangle$ in the plane of $\tilde \Omega$ and $\alpha$.
The black and white solid lines, corresponding to Eqs.~\eqref{eq:omega_avg_valley} and \eqref{eq:omega_avg_peak}, respectively, well capture the numerically-extracted minima (red triangles)
and maxima (red circles) for $\alpha \le 1$.
At larger $\alpha$, the extrema move toward each other and eventually disappear, indicating the transition from a non-monotonic response to a monotonic curve with a broad shoulder. 
The same transition is also observed in \figrefp{fig:Figure_4}{d} for different values of $\epsilon$.
Nevertheless, the local-maximum values of $\langle \tilde \omega \rangle$ are approximately insensitive to $\epsilon$ over the range examined.


\section{Conclusion}

Our analysis demonstrates that viscoelasticity can qualitatively reshape the asynchronous rotation of an isolated magnetic particle in a Jeffreys fluid. 
Depending on the solvent-to-solution viscosity ratio and the polymer relaxation time, the particle's time-averaged angular velocity can depart substantially from the Newtonian prediction and develop local minima and maxima. 
This behavior reflects a competition among magnetic driving, delayed elastic response, and frequency-dependent viscous dissipation.
The non-monotonicity arises when polymeric memory relaxes only partially within a phase-slip cycle.
In this case, retained deformation makes the forward and backward trajectories inequivalent, while the crossover from zero-shear to solvent-controlled intracycle dissipation changes the frequency scaling of the net angular advance. 
An asymptotic expression capturing the non-monotonic dependence is then derived based on the small perturbation assumption.
Our work not only advances the physical understanding of magnetic particle dynamics in viscoelastic media, but also provides crucial insights for optimizing magnetic manipulation strategies in biomedical applications where viscoelastic environments are prevalent.


For future work, additional effects of the thermal fluctuations, boundary confinements, interparticle interactions, and multimode or nonlinear rheology should be taken into account.
They are sometimes essential in relevant applications of magnetic particles and may influence the asynchronous dynamics. 
Moreover, to validate our theoretical results, quantitative comparison with experiments should be conducted.


\begin{acknowledgments}

Z.Z. acknowledge Z. Xiong and Z. Liu for valuable discussions. This work was supported by the National Nature Science Foundation of China (Nos.\,12304247, T2325027, 12274448, and 12072010), the Fundamental Research Funds for the Central Universities (No.\,GW2025-ZY-04), and startup funds of Wenzhou Institute, University of Chinese Academy of Sciences (No.\,WIUCASQD2022004). 

\end{acknowledgments}







\bibliography{library}

@PREAMBLE{
 "\providecommand{\noopsort}[1]{}" 
 # "\providecommand{\singleletter}[1]{#1}%" 
}

@article{errill1969rheology,
  title={Rheology of blood},
  author={Errill, EW},
  journal={Physiological reviews},
  volume={49},
  number={4},
  pages={863--888},
  year={1969}
}

@book{barnes1989introduction,
  title={An introduction to rheology},
  author={Barnes, Howard A and Hutton, John Fletcher and Walters, Kenneth},
  volume={3},
  year={1989},
  publisher={Elsevier}
}

@article{helgesen1990nonlinear,
  title={Nonlinear phenomena in systems of magnetic holes},
  author={Helgesen, Geir and Pieranski, Piotr and Skjeltorp, Arne T},
  journal={Physical review letters},
  volume={64},
  number={12},
  pages={1425},
  year={1990},
  publisher={APS}
}

@article{mcnaughton2007physiochemical,
  title={Physiochemical microparticle sensors based on nonlinear magnetic oscillations},
  author={McNaughton, Brandon H and Agayan, Rodney R and Wang, Jane X and Kopelman, Raoul},
  journal={Sensors and Actuators B: Chemical},
  volume={121},
  number={1},
  pages={330--340},
  year={2007},
  publisher={Elsevier}
}

@article{tierno2008controlled,
  title={Controlled swimming in confined fluids of magnetically actuated colloidal rotors},
  author={Tierno, Pietro and Golestanian, Ramin and Pagonabarraga, Ignacio and Sagu{\'e}s, Francesc},
  journal={Physical review letters},
  volume={101},
  number={21},
  pages={218304},
  year={2008},
  publisher={APS}
}

@article{tierno2009overdamped,
  title={Overdamped dynamics of paramagnetic ellipsoids in a precessing magnetic field},
  author={Tierno, Pietro and Claret, Josep and Sagu{\'e}s, Francesc and C{\=e}bers, Andrejs},
  journal={Physical Review E—Statistical, Nonlinear, and Soft Matter Physics},
  volume={79},
  number={2},
  pages={021501},
  year={2009},
  publisher={APS}
}

@article{frka2011dynamics,
  title={Dynamics of paramagnetic nanostructured rods under rotating field},
  author={Frka-Petesic, Bruno and Erglis, Kaspars and Berret, Jean-Francois and Cebers, Andrejs and Dupuis, Vincent and Fresnais, J{\'e}r{\^o}me and Sandre, O and Perzynski, R},
  journal={Journal of Magnetism and Magnetic Materials},
  volume={323},
  number={10},
  pages={1309--1313},
  year={2011},
  publisher={Elsevier}
}

@article{sinn2011asynchronous,
  title={Asynchronous magnetic bead rotation (AMBR) biosensor in microfluidic droplets for rapid bacterial growth and susceptibility measurements},
  author={Sinn, Irene and Kinnunen, Paivo and Albertson, Theodore and McNaughton, Brandon H and Newton, Duane W and Burns, Mark A and Kopelman, Raoul},
  journal={Lab on a Chip},
  volume={11},
  number={15},
  pages={2604--2611},
  year={2011},
  publisher={Royal Society of Chemistry}
}

@book{fung2013biomechanics,
  title={Biomechanics: mechanical properties of living tissues},
  author={Fung, Yuan-cheng},
  year={2013},
  publisher={Springer Science \& Business Media}
}

@article{chevry2013magnetic,
  title={Magnetic wire-based sensors for the microrheology of complex fluids},
  author={Chevry, L and Sampathkumar, NK and Cebers, A and Berret, J-F},
  journal={Physical Review E—Statistical, Nonlinear, and Soft Matter Physics},
  volume={88},
  number={6},
  pages={062306},
  year={2013},
  publisher={APS}
}

@article{tokarev2013probing,
  title={Probing viscosity of nanoliter droplets of butterfly saliva by magnetic rotational spectroscopy},
  author={Tokarev, Alexander and Kaufman, Bethany and Gu, Yu and Andrukh, Taras and Adler, Peter H and Kornev, Konstantin G},
  journal={Applied Physics Letters},
  volume={102},
  number={3},
  year={2013},
  publisher={AIP Publishing}
}

@article{tierno2014recent,
  title={Recent advances in anisotropic magnetic colloids: realization, assembly and applications},
  author={Tierno, Pietro},
  journal={Physical chemistry chemical physics},
  volume={16},
  number={43},
  pages={23515--23528},
  year={2014},
  publisher={Royal Society of Chemistry}
}

@article{haapanen2015observation,
  title={Observation of viscoelastic solutions with ferromagnetic stirrers},
  author={Haapanen, Karoliina and Kinnunen, Paivo and Czajkowski, Jakub and Lahti, Anni and Illikainen, Mirja},
  journal={Sensors and Actuators A: Physical},
  volume={236},
  pages={309--314},
  year={2015},
  publisher={Elsevier}
}

@article{berret2016local,
  title={Local viscoelasticity of living cells measured by rotational magnetic spectroscopy},
  author={Berret, J-F},
  journal={Nature communications},
  volume={7},
  number={1},
  pages={10134},
  year={2016},
  publisher={Nature Publishing Group UK London}
}

@article{loosli2016viscoelasticity,
  title={Viscoelasticity of model surfactant solutions determined by magnetic rotation spectroscopy},
  author={Loosli, F and Najm, M and Berret, J-F},
  journal={Colloids and Surfaces A: Physicochemical and Engineering Aspects},
  volume={510},
  pages={143--149},
  year={2016},
  publisher={Elsevier}
}

@article{egolf2016hyperthermia,
  title={Hyperthermia with rotating magnetic nanowires inducing heat into tumor by fluid friction},
  author={Egolf, Peter W and Shamsudhin, Naveen and Pan{\'e}, Salvador and Vuarnoz, Didier and Pokki, Juho and Pawlowski, Anne-Gabrielle and Tsague, Paulin and De Marco, Bastien and Bovy, William and Tucev, Sinisa and others},
  journal={Journal of Applied Physics},
  volume={120},
  number={6},
  year={2016},
  publisher={AIP Publishing}
}

@article{erb2016actuating,
  title={Actuating soft matter with magnetic torque},
  author={Erb, Randall M and Martin, Joshua J and Soheilian, Rasam and Pan, Chunzhou and Barber, Jabulani R},
  journal={Advanced Functional Materials},
  volume={26},
  number={22},
  pages={3859--3880},
  year={2016},
  publisher={Wiley Online Library}
}

@article{ponce2018validity,
  title={On the validity of the Jeffreys (Oldroyd-B) model to describe the oscillations of a viscoelastic pendant drop},
  author={Ponce-Torres, A and Acero, AJ and Herrada, MA and Montanero, JM},
  journal={Journal of Non-Newtonian Fluid Mechanics},
  volume={260},
  pages={69--75},
  year={2018},
  publisher={Elsevier}
}

@article{moerland2019rotating,
  title={Rotating magnetic particles for lab-on-chip applications--a comprehensive review},
  author={Moerland, CP and Van IJzendoorn, LJ and Prins, MWJ},
  journal={Lab on a Chip},
  volume={19},
  number={6},
  pages={919--933},
  year={2019},
  publisher={Royal Society of Chemistry}
}

@article{soni2019odd,
  title={The odd free surface flows of a colloidal chiral fluid},
  author={Soni, Vishal and Bililign, Ephraim S and Magkiriadou, Sofia and Sacanna, Stefano and Bartolo, Denis and Shelley, Michael J and Irvine, William TM},
  journal={Nature physics},
  volume={15},
  number={11},
  pages={1188--1194},
  year={2019},
  publisher={Nature Publishing Group UK London}
}

@article{xie2019reconfigurable,
  title={Reconfigurable magnetic microrobot swarm: Multimode transformation, locomotion, and manipulation},
  author={Xie, Hui and Sun, Mengmeng and Fan, Xinjian and Lin, Zhihua and Chen, Weinan and Wang, Lei and Dong, Lixin and He, Qiang},
  journal={Science robotics},
  volume={4},
  number={28},
  pages={eaav8006},
  year={2019},
  publisher={American Association for the Advancement of Science}
}

@article{yang2021motion,
  title={Motion control in magnetic microrobotics: From individual and multiple robots to swarms},
  author={Yang, Lidong and Zhang, Li},
  journal={Annual Review of Control, Robotics, and Autonomous Systems},
  volume={4},
  number={1},
  pages={509--534},
  year={2021},
  publisher={Annual Reviews}
}

@article{zhou2021magnetically,
  title={Magnetically driven micro and nanorobots},
  author={Zhou, Huaijuan and Mayorga-Martinez, Carmen C and Pan{\'e}, Salvador and Zhang, Li and Pumera, Martin},
  journal={Chemical Reviews},
  volume={121},
  number={8},
  pages={4999--5041},
  year={2021},
  publisher={ACS Publications}
}

@article{massana2021arrested,
  title={Arrested phase separation in chiral fluids of colloidal spinners},
  author={Massana-Cid, Helena and Levis, Demian and Hern{\'a}ndez, Ra{\'u}l Josu{\'e} Hern{\'a}ndez and Pagonabarraga, Ignacio and Tierno, Pietro},
  journal={Physical Review Research},
  volume={3},
  number={4},
  pages={L042021},
  year={2021},
  publisher={APS}
}

@article{cai2023magnetic,
  title={Magnetic bead manipulation in microfluidic chips for biological application},
  author={Cai, Gaozhe and Yang, Zixin and Chen, Yu-Cheng and Huang, Yaru and Liang, Lijuan and Feng, Shilun and Zhao, Jianlong},
  journal={Cyborg and bionic systems},
  volume={4},
  pages={0023},
  year={2023},
  publisher={AAAS}
}

@article{wang2023reconfigurable,
  title={Reconfigurable vortex-like paramagnetic nanoparticle swarm with upstream motility and high body-length ratio velocity},
  author={Wang, Luyao and Gao, Han and Sun, Hongyan and Ji, Yiming and Song, Li and Jia, Lina and Wang, Chutian and Li, Chan and Zhang, Deyuan and Xu, Ye and others},
  journal={Research},
  volume={6},
  pages={0088},
  year={2023},
  publisher={AAAS}
}

@article{barrera2024magnetic,
  title={Magnetic nanoparticle hyperthermia enhanced by a rotating field},
  author={Barrera, Gabriele and Allia, Paolo and Tiberto, Paola},
  journal={Physical Review Applied},
  volume={21},
  number={6},
  pages={064037},
  year={2024},
  publisher={APS}
}

@article{chen2025self,
  title={Self-propulsion, flocking and chiral active phases from particles spinning at intermediate Reynolds numbers},
  author={Chen, Panyu and Weady, Scott and Atis, Severine and Matsuzawa, Takumi and Shelley, Michael J and Irvine, William TM},
  journal={Nature Physics},
  volume={21},
  number={1},
  pages={146--154},
  year={2025},
  publisher={Nature Publishing Group UK London}
}

@article{luo2025flocking,
  title={Flocking Phase Separation in Inertial Active Matter},
  author={Luo, Nan and Li, Longfei and Yang, Mingcheng and Peng, Yi},
  journal={Physical Review Letters},
  volume={135},
  number={17},
  pages={178301},
  year={2025},
  publisher={APS}
}

@article{landers2025clinically,
  title={Clinically ready magnetic microrobots for targeted therapies},
  author={Landers, Fabian C and Hertle, Lukas and Pustovalov, Vitaly and Sivakumaran, Derick and Oral, Cagatay M and Brinkmann, Oliver and Meiners, Kirstin and Theiler, Pascal and Gantenbein, Valentin and Veciana, Andrea and others},
  journal={Science},
  volume={390},
  number={6774},
  pages={710--715},
  year={2025},
  publisher={American Association for the Advancement of Science}
}

@article{gao2025response,
  title={Response of a magnetic particle to rotating magnetic field in viscoelastic fluid},
  author={Gao, Han and Zhao, Zhiyuan and Doi, Masao and Xu, Ye},
  journal={Physical Review E},
  volume={112},
  number={5},
  pages={055417},
  year={2025},
  publisher={APS}
}
\end{document}